\documentclass[11pt]{article}
\usepackage{amssymb,amsmath,amsthm,graphicx,ulem}
\usepackage{graphicx}
\usepackage{epsfig}
\usepackage{bm} 
\usepackage{amsfonts,amssymb,amsmath}
\usepackage{latexsym}
\usepackage[english]{babel}
\usepackage{subfigure}

\def\beq{\begin{equation}}
\def\eeq{\end{equation}}
\def\bea{\begin{eqnarray}}
\def\eea{\end{eqnarray}}

\newcommand{\eqn}{\begin{eqnarray}}
\newcommand{\eeqn}{\end{eqnarray}}
\newcommand{\arr}{\begin{eqnarray*}}
\newcommand{\earr}{\end{eqnarray*}}

\newcommand{\ie}{{i.e.}\ } 

\newcommand{\lp}{\left(}
\newcommand{\rp}{\right)}

\def\ii{\textrm{i}}

\begin{document}

\setlength{\unitlength}{1mm}

\thispagestyle{empty} 
 \vspace*{2cm}

\begin{center}
{\bf \LARGE  Ultraspinning instability: the missing link}\\

\vspace*{2.5cm}

 {\bf \'Oscar J.~C.~Dias$^{a}\,$},
{\bf Ricardo Monteiro$^{b}\,$}, {\bf Jorge E.~Santos$^{c}\,$}
 \vspace*{0.5cm}

{\it $^a\,$ DAMTP, Centre for Mathematical Sciences, University of Cambridge,\\
Wilberforce Road, Cambridge CB3 0WA, United Kingdom}\\[.3em]
{\it $^b\,$ The Niels Bohr International Academy, The Niels Bohr Institute, \\
Blegdamsvej 17, DK-2100 Copenhagen, Denmark}\\[.3em]
{\it $^c\,$ Department of Physics, UCSB, Santa Barbara, CA 93106, USA}\\[.3em]

\vspace*{0.5cm} {\tt O.Dias@damtp.cam.ac.uk, monteiro@nbi.dk, jss55@physics.ucsb.edu}

\end{center}

\begin{abstract}

We study linearized perturbations of Myers-Perry black holes in $d=7$, with two of the three angular momenta set to be equal, and show that instabilities always appear before extremality. Analogous results are expected for all higher odd $d$. We determine numerically the stationary perturbations that mark the onset of instability for the modes that preserve the isometries of the background. The onset is continuously connected between the previously studied sectors of solutions with a single angular momentum and solutions with all angular momenta equal. This shows that the near-extremality instabilities are of the same nature as the ultraspinning instability of $d\geq6$ singly-spinning solutions, for which the angular momentum is unbounded. Our results raise the question of whether there are any extremal Myers-Perry black holes which are stable in $d\geq6$.

\end{abstract}

\noindent


\vfill \setcounter{page}{0} \setcounter{footnote}{0}
\newpage

\tableofcontents



\setcounter{equation}{0}
\section{Introduction}

In recent years, we have acquired a greater understanding of the phase space of higher-dimensional black holes. The main motivation for this research is that string theory predicts the existence of higher spacetime dimensions. Other scenarios with extra dimensions have been proposed in a phenomenological approach, independently of string theory. It is therefore crucial to understand what kinds of objects exist in higher-dimensional spacetimes. Black holes are the simplest gravitational objects, and they are essential to the study of fundamental theories. They are also the main candidates to signal the possible existence of extra dimensions in high-energy collisions.

If higher-dimensional black holes can be produced in our universe, which are the relevant solutions? In four spacetime dimensions, the answer is simple: there is a unique vacuum solution, the Kerr black hole, which is (expected to be) stable. However, uniqueness fails in higher dimensions \cite{Emparan:2008eg}. The higher-dimensional version of the Kerr black hole, the Myers-Perry (MP) solution, has long been known \cite{Myers:1986un}. It was the discovery of the black ring by Emparan and Reall \cite{Emparan:2001wn} which showed that several rotating solutions may exist with the same asymptotic charges (mass and angular momenta), and that different topologies of the event horizon are possible. Another important feature is the possibility of black hole solutions with disconnected horizons, such as the black saturn \cite{Elvang:2007rd} and concentric rings \cite{Iguchi:2007is,Evslin:2007fv}, among others; this is believed to be impossible in four dimensions (in vacuum). Actually, only the MP solution is known exactly for $d>5$. The new solutions were obtained in $d=5$, where powerful analytical techniques are available, but they are expected to exist in any higher dimensions. They have in fact been constructed approximately in a certain regime of high rotation \cite{Emparan:2007wm}, in what has recently been put into a systematic formalism -- the blackfold approach \cite{Emparan:2009cs,Emparan:2009at,Emparan:2009vd}.

An obvious question is whether these solutions are stable. There is one property of higher-dimensional black holes which is fundamental to this question: the event horizon can have very different length scales. In four dimensions, the horizon of the Kerr black hole is a deformed sphere, and the deformation increases with the rotation. However, that deformation can never be too pronounced, because the rotation is bounded by extremality. This type of bound does not exist for some higher-dimensional solutions, so that regimes exist where there is a hierarchy between the scales set by the angular momenta and by the mass. When the mass scale is much larger, one expects a similar behaviour as in four dimensions. When an angular momentum scale is much larger, new features arise which are characteristic of higher dimensions.

The approximation of large rotation, which is the basis of the blackfold approach mentioned above, was explored by Emparan and Myers to argue that singly-spinning MP black holes are unstable if rotating too rapidly \cite{Emparan:2003sy}. Singly-spinning MP solutions have the isometry group ${\mathbb R}\times U(1)\times SO(d-3)$, which is an enhancement from the isometry group ${\mathbb R} \times U(1)^n$ of generic MP black holes ($n\equiv \lfloor (d-1)/2 \rfloor$ is the number of independently defined angular momenta). The argument for instability is that, for very high rotation, the solution acquires a disk-like horizon and resembles a black brane locally. It has been shown by Gregory and Laflamme \cite{Gregory:1993vy} that black branes are unstable, so these ultraspinning MP black holes should become unstable too. Ref.~\cite{Emparan:2003sy} also pointed out that, for perturbations which do not break the $U(1)$ symmetry (which is the angle associated to the rotation of the black hole), the onset of the instability should be a stationary mode which signals the bifurcation to a new black hole family. This idea was elaborated on Ref.~\cite{Emparan:2007wm}, which conjectured that there is a harmonic-type structure of ultraspinning instabilities, and that the new black hole solutions appearing at the associated bifurcations interpolate between the MP family and the black ring, the black saturn, concentric rings, etc..

Refs.~\cite{Dias:2009iu,Dias:2010maa} provided strong evidence for this picture of the phase space of singly-spinning black holes. The critical rotations and the stationary onset modes were determined numerically. Moreover, those modes were found to be consistent with the expected horizon shape of the new families, if they are to connect to the black ring, black saturn, etc.. Ref.~\cite{Dias:2009iu} put forward a conjecture about the appearance of these instabilities beyond the singly-spinning sector. It says that instabilities whose onset is a stationary mode can only appear if the black hole has two local thermodynamic instabilities. Note that this is a necessary, but not sufficient condition for the appearance of such an instability. We refer to this conjecture as the ultraspinning conjecture. It seems strange to relate classical and local thermodynamic stability for non-extended objects (as opposed to the Gubser-Mitra conjecture for black branes \cite{Gubser:2000ec}). The point is that these instabilities are part of the same harmonic structure. The two lowest harmonics are associated with the asymptotic charges and with the local thermodynamic instabilities, while the higher harmonics are associated with modes which cannot change the asymptotic charges, and are associated with the classical instabilities.

The goal of the ultraspinning conjecture is to provide a guidance for the appearance of instabilities. The great advantage is that it does not rely on hierarchies between the mass and the angular momenta scales, telling us something about the intermediate regime where the scales are of the same order. The conjecture was the motivation for the work of Ref.~\cite{Dias:2010eu}, which studied the stability of cohomogeneity-1 MP black holes. This sector consists of odd-dimensional solutions which have all $n$ angular momenta turned on, but equal. There is an enhancement of symmetry to the isometry group ${\mathbb R}\times U(N+1)$, where $d=2N+3$; this leads to a line element which depends non-trivially on the radial coordinate only, i.e. which has cohomogeneity-1 \cite{Kunduri:2006qa}. The regime of large rotation, in the sense that the angular momentum is much larger than the scale set by the mass, does not exist in this sector, because the rotation is bounded by extremality. However, the ultraspinning conjecture allows for a small parameter region close to extremality where an instability is possible, in $d\geq7$. The $d=5$ solution is stable \cite{Murata:2008yx}.

An instability in the cohomogeneity-1 sector was indeed found in $d=9$ \cite{Dias:2010eu}. Moreover, due to the symmetry of these solutions, it was possible to determine the growth rate of the unstable modes, which increases with rotation, while the works on the singly-spinning sector could only find the stationary onset. This was an explicit confirmation that we are dealing with actual instabilities. Ref.~\cite{Dias:2010eu} failed to find that the instability appears already in $d=7$. It is so close to extremality, $(J_{\textrm{inst}}-J_{\textrm{ext}})/J_{\textrm{ext}} \sim 10^{-5}$, that it did not show up in the original analysis. It was revealed in a reanalysis performed by one of us (JES), by request from the authors of Ref.~\cite{Durkee:2010ea}, where it is argued that the extremal black hole is unstable. The instability extends to a small parameter region close to extremality, as the numerics confirmed. Ref.~\cite{Durkee:2010ea} makes use of an extension of the Newman-Penrose formalism to higher dimensions \cite{Durkee:2010qu}, and studies the stability of extremal solutions by analysing their near-horizon geometry. It is conjectured there that the existence of an instability of the near-horizon geometry implies the existence of an instability of the full black hole geometry, if the near-horizon perturbations satisfy a certain symmetry.

The story seems satisfactory so far, and the ultraspinning conjecture has been shown to hold also in the singly-spinning asymptotically AdS case \cite{Dias:2010gk}, where a large rotation limit analogous to the asymptotically flat case exists \cite{Caldarelli:2008pz,Armas:2010hz}. However, an important question remains. Why should the instability in the cohomogeneity-1 sector be related to the instability in the singly-spinning sector? While in the latter case there is the convincing argument of Emparan and Myers, no such clear geometric understanding of the instability exists for the cohomogeneity-1 sector. One can see the ultraspinning conjecture as extending the notion of brane-like behaviour, and thus the argument of Emparan and Myers. However, it is important to see explicitly how these two sectors fit together in the stability picture of the MP family. This provides the motivation for the present paper, where we will study a MP sub-family which includes the cohomogeneity-1 sector and the singly-spinning sector as particular cases.

We will consider MP black holes in an odd number of spacetime dimensions, which have all but one of the $n$ angular momenta equal, so that we have effectively two unequal spins. The isometry group of this MP sub-family is ${\mathbb R}\times U(1)\times U(N)$ (recall that $d=2N+3$). We particularize our study to $d=7$, and restrict ourselves, for simplicity, to perturbations which preserve the isometries of the background solutions. As a consequence, the perturbed geometries will also have two unequal spins. This restriction still allows us to find the missing link between the singly-spinning and the cohomogeneity-1 sectors. One of the main results of this paper is that the onset of the ultraspinning instability is continuously connected between those two sectors in the MP parameter space. This shows that the instabilities are indeed of the same nature.

The most important outcome of our analysis, however, goes beyond the initial motivation. We find that all near-extremal solutions studied here are unstable. This raises the question of whether there is any extremal MP black hole which is stable in $d\geq 6$, i.e. when the parameter space admits unbounded sectors. 

Before proceeding, let us point out that fastly rotating MP black holes can also be unstable against bar-mode perturbations that break the rotation-generating symmetry of the background. This is the case at least for singly-spinning MP black holes \cite{Shibata:2009ad,Shibata:2010wz}. We emphasize that the ultraspinning conjecture only applies to instabilities whose onset is a stationary mode, which is not the case for the bar-mode perturbations. Therefore, the bar-mode instability avoids the conjectured bound, and even occurs in $d=5$, but it is not associated with a bifurcation to a new family of stationary black holes. It would be important to determine how this type of instability manifests itself beyong the singly-spinning sector. In $d=5$, Ref.~\cite{Murata:2008yx} found no evidence of instability for equal spins.

This paper is organised as follows. In Section~\ref{sec:MPbackground}, we present the MP sub-family with two unequal spins, and show how the ultraspinning conjecture applies to it. In Section~\ref{sec:PerturbationProblem}, we describe the linear perturbation problem that is solved numerically. We conclude in section~\ref{sec:results}, with the discussion of the results.

\setcounter{equation}{0}
\section{Myers-Perry black holes with two unequal spins}
\label{sec:MPbackground}

\subsection{Solution}
\label{subsec:MP}

The Myers-Perry (MP) black hole \cite{Myers:1986un} is an asymptotically flat solution to the vacuum Einstein equations, which generalizes the Kerr solution to higher dimensions. This family of black holes can be parameterized by the asymptotic charges of the spacetime, the mass $M$ and the $n\equiv \lfloor (d-1)/2 \rfloor$ angular momenta $J_i$; or, equivalently, by the mass-radius $r_M$ and the rotation parameters $a_i$, to be defined below.

A generic MP solution has isometry group $\mathbb{R} \times U(1)^n$, where $\mathbb{R}$ corresponds to time translations and each $U(1)$ describes one of the $n$ rotational isometries. However, there is an enhancement of symmetry when some of the angular momenta coincide. We will be interested in the sub-family of MP black holes that has $n-1$ equal rotation parameters, denoted by $a$, while the remaining rotation parameter, denoted by $b$, is independent. In odd-dimensional spacetimes, which we will focus on, such solutions have the isometry group $\mathbb{R} \times U(1) \times U(N)$, where $d=2N+3$. This family interpolates between the singly-spinning MP black hole and the equal angular momenta MP black hole.
The former is obtained by setting $a=0$, and has isometry group $\mathbb{R}\times U(1)\times SO(d-3)$. The latter is obtained by setting $a=b$, and (for odd $d$) has isometry group $\mathbb{R} \times U(N+1)$; due to this enhancement of symmetry, this solution is cohomogeneity-1, \ie it depends non-trivially on the radial coordinate only.

For concreteness, we study here the MP sub-family with two unequal spins in $d=7$. (In $d=5$, there is no instability of the type which we will consider.) We then have $a_1=a_2\equiv a$ and $a_3\equiv  b$. The line element of this solution, which can be derived from \cite{Oota:2008uj}, is
\begin{eqnarray}\label{eq:MPab}
&& \hspace{-0.7cm}ds^2= \frac{r^2+v^2}{X(r)}\,dr^2+\frac{r^2+v^2}{Y(v)}\,dv^2+\frac{(r^2+a^2)(a^2-v^2)}{a^2-b^2}\,d s^2_{\mathbb{CP}^1} \nonumber\\
&& \hspace{0.2cm} -\frac{X(r)}{r^2+v^2}\left[dt+\frac{a(a^2-v^2)}{a^2-b^2} (d \psi_1+\mathbb{A}_1)\right]^2 +\frac{Y(v)}{r^2+v^2}\frac{a^2(r^2+a^2)^2}{(a^2-b^2)^2}\left[(d \psi_1+\mathbb{A}_1)+\frac{a^2-b^2}{a(r^2+a^2)}\,dt\right]^2 \nonumber \\
&& \hspace{0.2cm}+r^2v^2a^2b^2\left[\frac{d\psi_2}{a^2 b}-\frac{(r^2+a^2)(a^2-v^2)}{a(a^2-b^2)r^2v^2}(d \psi_1+\mathbb{A}_1)-\frac{dt}{r^2v^2}\right]^2,
\end{eqnarray}
where we defined the functions
\begin{equation}
X(r) = \frac{(r^2+a^2)(r^2+b^2)}{r^2}-\frac{r_M^4}{r^2+a^2},\qquad Y(v) = -\frac{(a^2-v^2)(b^2-v^2)}{v^2},
\end{equation}
and
\begin{equation}
ds^2_{\mathbb{CP}_1}=\frac{1}{4}\left( d\theta^2+ \sin^2\theta \,d\varphi^2\right),\qquad  \mathbb{A}_1 =\frac{1}{2}\,\cos\theta \,d\varphi,
\end{equation}
are, respectively, the line element of $\mathbb{CP}_1\cong S^2$ and the associated  K\"ahler 1-form  $\mathbb{A}_1$  (related to the K\"ahler 2-form $\mathbb{J}_1$ by $d\mathbb{A}_1=2\mathbb{J}_1$). 
The time $t$ and radial $r$ coordinates have the standard range, $-\infty<t<+\infty,\,0\leq r<\infty$, and the angular coordinates $v,\psi_1,\psi_2,\theta,\phi$ range in the intervals 
 \begin{equation}\label{eq:coordRange}
\hbox{min}\{|a|,|b|\} \leq v\leq \hbox{max}\{|a|,|b|\} \,,\qquad 0\leq\psi_1\,,\psi_2<2\pi\,,\qquad  0\leq\theta\leq\pi\,,\qquad 0\leq\varphi<2\pi.
\end{equation}

The mass $M$ and the two unequal angular momenta $J^{(\psi_1)}$ and $J^{(\psi_2)}$ are given by
\begin{equation}
 M=\frac{5 \pi ^2r_M^4}{16}, \qquad J^{(\psi_1)}=\frac{\pi ^2r_M^4\,a}{4}, \qquad J^{(\psi_2)}=\frac{\pi ^2r_M^4\,b}{8}.
\end{equation}
Note that a $d=7$ spacetime has three asymptotically defined angular momenta ($n=3$). However, since two of them are equal, it is more convenient to define from the line element \eqref{eq:MPab} two angular momenta, with respect to $\partial_{\psi_1}$ and $\partial_{\psi_2}$.

The event horizon is located at $r=r_+$, where $r_+$ denotes the largest real root of $X(r)=0$, and it is a Killing horizon of $\xi = \partial_t+\Omega_H^{(\psi_1)} \partial_{\psi_1}+\Omega_H^{(\psi_2)} \partial_{\psi_2}$, where the angular velocities of the horizon are given by
\begin{equation}\label{eq:angVel}
\Omega_H^{(\psi_1)}=\frac{b^2-a^2}{a \left(r_+^2+a^2\right)}\,,\qquad \Omega_H^{(\psi_2)}=\frac{b}{r_+^2}\,.
\end{equation}
Finally, the Hawking temperature $T_H$ and the Bekenstein-Hawking entropy $S$ of the solution are given by
\begin{equation}
T_H=\frac{r_+}{2 \pi }\left(\frac{r_+^2-a^2}{r_+^2\left(r_+^2+a^2\right)}+\frac{1}{r_+^2+b^2}\right), \qquad
S=\frac{\pi ^3r_M^4 r_+}{4}. 
\end{equation}
In terms of these quantities, the first law of thermodynamics takes the form
\begin{equation}
dM=T_H dS + \Omega_H^{(\psi_1)} dJ^{(\psi_1)} + \Omega_H^{(\psi_2)} dJ^{(\psi_2)}.
\end{equation}
Since we defined the angular velocities/momenta with respect to $\partial_{\psi_1}$ and $\partial_{\psi_2}$, this form of the first law does not allow for variations of parameters in the MP family which go beyond the two unequal spins sub-family; that is, this expression restricts the variations to be within the line element \eqref{eq:MPab}. This is important for us, because we will make use of the thermodynamics to indicate the possibility of classical instabilities of the black hole, and we will consider only perturbations preserving the equality between two of the three angular momenta. 

We presented here the general two unequal spins solution. Since our goal is to connect previous work on the singly-spinning sector and on the equal spins sector, we will analyze those particular limits in detail. The coordinate transformations
\begin{equation}\label{eq:TranfToEqual}
t=T-a \psi\,,\qquad r=\sqrt{\rho^2-a^2}\,,\qquad v=a+(b-a)\,\frac{R^2}{1+R^2}\,,\qquad 
\psi _1=\frac{1}{2}\Psi\,,\qquad \psi_2=\frac{b}{a}\,\psi\,,
\end{equation}
followed by the limit $b\to a$ take the MP solution \eqref{eq:MPab} into the equal angular momenta MP black hole \cite{Kunduri:2006qa}:
\begin{eqnarray} \label{eq:MPequalJ}
&&\hspace{-1.5cm}ds^2 = -f(\rho)\,g(\rho)\,dT^2 +\frac{d\rho^2}{f(\rho)} + \rho^2 h(\rho)\,[d\psi +\mathbb{A}_2  - \Omega(\rho)\,dT]^2 + \rho^2 ds^2_{\mathbb{CP}^2}, \qquad \hbox{where}  \nonumber \\
&& \hspace{-0.8cm} f(\rho)=1-\frac{r_M^4}{\rho^4}+\frac{r_M^4 a^2}{\rho^6}\,,\qquad g(\rho)=\frac{1}{h(\rho)}\,,\qquad h(\rho)=1+\frac{r_M^4 a^2}{\rho^6}\,,\qquad \Omega(\rho) =\frac{r_M^4 a}{\rho^6 h(\rho)}, \\
&& \hspace{-0.8cm}  ds^2_{\mathbb{CP}^2} = \frac{dR^2}{\left( 1+R^2\right)^2}
  +\frac{R^2}{\left( 1+R^2\right)^2}\,\left( d\Psi+\mathbb{A}_{1} \right)^2 +
\frac{R^2}{ 1+R^2}\, ds^2_{\mathbb{CP}^1}\,,\qquad  \mathbb{A}_2=\frac{R^2}{1+R^2}\left( d\Psi +
  \mathbb{A}_{1} \right).\nonumber
\end{eqnarray}
Here, $ds^2_{\mathbb{CP}^2}$ is the  Fubini-Study metric on $\mathbb{CP}^2$, while $\mathbb{A}_2$ is the associated K\"ahler 1-form.
In this solution, surfaces of constant $T$ and $\rho$ have the geometry of a homogeneously squashed $S^{5}$, written as an $S^1$ fibred over $\mathbb{CP}^{2}$. The fibre is parameterized by the coordinate $\psi$, which has period $2\pi$. The largest real root of $f(\rho)=0$ gives the location of the event horizon, $\rho=\rho_+$, which is a Killing horizon of $\xi = \partial_t+\Omega_H\partial_{\psi}$ with $\Omega_H =  a/\rho_+^2$. The angular momentum of this black hole is bounded, being maximal for a regular extremal solution (the temperature vanishes but not the entropy).

On the other hand, the coordinate transformations
\begin{equation}\label{eq:TranfToSingle}
t = T - b\, \phi\,,\qquad r = r \,,\qquad v = b\, \cos \theta \,,\qquad \psi_1 = \Psi + \frac{b}{a}\, \phi\,,\qquad  \psi_2 = \phi\,,
\end{equation}
followed by the limit $a\to 0$, take the MP solution \eqref{eq:MPab} into the singly-spinning MP black hole  \cite{Myers:1986un}:
\begin{eqnarray}\label{eq:MPsingleJ}
&&\hspace{-1.4cm}ds^2 = -\frac{\Delta}{\Sigma}\left( dT-b\sin^2\theta
\,d\phi \right)^2 + \frac{\sin^2\theta}{\Sigma}\left[
(r^2+b^2)d\phi-b\,dT\right]^2
+\frac{\Sigma}{\Delta}\,dr^2+
\Sigma\,d\theta^2 +r^2\cos^2\theta\, d\Omega^2_{S^3},  \nonumber \\
&& \hspace{-0.8cm} \Delta(r)=r^2+b^2-\frac{r_M^{4}}{r^2}\,,\qquad  \Sigma(r,\theta)=r^2+b^2\cos^2\theta\,, \qquad d\Omega^2_{S^3}= (d \Psi+\mathbb{A}_1)^2+ ds^2_{\mathbb{CP}^1}. 
\end{eqnarray}
Note that $d\Omega^2_{S^3}$ is the line element of a unit $S^3$, here written as a Hopf fibration of $S^1$ over $\mathbb{CP}^1\cong S^2$. The largest real root of $\Delta(r)=0$ gives the location of the event horizon, $r=r_+$, which is a Killing horizon of $\xi = \partial_t+\Omega_H\partial_{\phi}$ with  $\Omega_H=\frac{b}{r_+^2 + b^2}$. For the singly-spinning MP black hole (in $d>5$), there is no bound on the angular momentum.

Let us also mention the case $b=0$. There is an extremality bound on the rotation parameter $a$ of such a solution, but the limit is singular (both the temperature and the entropy vanish). In summary, the rotation parameters of the two unequal spins MP sub-family are bounded by regular extremal solutions, except for $a=0$ (where no extremal limit exists) and for $b=0$ (where an extremal limit exists, but is singular).

\subsection{Thermodynamic zero-modes and the ultraspinning regime \label{subsec:ultrathermo}}

The {\it ultraspinning conjecture} proposed in Refs.~\cite{Dias:2009iu,Dias:2010eu} says that classical instabilities of vacuum black holes whose onset is a stationary and axisymmetric mode can only appear in a parameter space region (called the ultraspinning region) determined by the existence of two local thermodynamic instabilities. By axisymmetric we mean that the symmetry generated by the Killing vector field $\Omega_i m_i$ is preserved; here, the $m_i$'s denote the rotational Killing vector fields and the $\Omega_i$'s denote the associated angular velocities of the horizon. If axisymmetry is broken, then stationarity is not possible, by virtue of the rigidity theorem \cite{Hollands:2006rj,Moncrief:2008mr}. In this section, we review this conjecture, which is essential to interpret our results.

The condition for local thermodynamic stability is the positivity of the thermodynamic Hessian
\begin{equation}
\label{thermoHessian} -S_{\alpha\beta}\equiv
-\,\frac{\partial^2{S(x_\gamma)}}{{\partial x_\alpha}{\partial
x_\beta}} \,, \qquad \; x_\alpha=(M,J_i)\,,
\end{equation}
where $S$ is the black hole entropy, $M$ its mass and $J_i$ denote all the possible angular momenta. Using only the  Smarr relation and the first law, it was shown in Ref.~\cite{Dias:2010eu} that $-S_{\alpha\beta}$ possesses at least one negative eigenvalue for any asymptotically flat vacuum black hole. Hence all such black holes are locally thermodynamically unstable. The ultraspinning region is defined as the parameter space region where there are at least two negative eigenvalues of the thermodynamic Hessian.

Let us elucidate the conjecture by considering MP black holes. For small angular momenta, there is a single negative eigenvalue of the Hessian \eqref{thermoHessian}, which is the one continuously connected to the negative specific heat of the Schwarzschild black hole. However, for $d>5$, additional eigenvalues vanish and then become negative as the rotation increases; the vanishing of the first of these signals the {\it ultraspinning surface}. So the ultraspinning surface defines a ``small angular momenta" region in the MP parameter space, where a single eigenvalue is negative, and no instability (whose onset is a stationary mode) is allowed. Beyond the ultraspinning surface, we have the ultraspinning region, where instabilities are allowed, but not required.

Whenever the thermodynamic Hessian has a vanishing eigenvalue, the corresponding eigenvector $(\delta M, \delta J_i)$ represents a perturbation of the black hole (within the MP family, in our example) which changes its mass and angular momenta, but preserves its temperature and angular velocities \cite{Dias:2010eu}. This type of perturbation is associated with the $\ell=0$ and $\ell=1$ ``harmonics" of the black hole; these contribute to the asymptotic charges. What then corresponds to the $\ell\geq2$ harmonics? If such zero-mode perturbations were allowed, they would change the solution but not the asymptotic charges, as they decay too quickly with large radius. They would then indicate a bifurcation to a new black hole family which can have the same asymptotic charges of the unperturbed solution. In the MP case, such new families of black holes where conjectured to exist (in $d>5$) and to provide a connection in parameter space between the MP family and other families of black holes, such as black rings, black saturns, etc. \cite{Emparan:2007wm,Emparan:2010sx}. Moreover, the MP black hole is expected to become unstable for rotations larger than that critical value of bifurcation \cite{Emparan:2007wm}. The ultraspinning conjecture reflects the expectation that a zero-mode associated with the $\ell\geq2$ harmonics, and thus with the onset of the classical instability, appears only for rotations larger than the $\ell\geq1$ harmonics.

In fact, there is a precise sense in which the harmonics $\ell\geq2$ can acquire a negative eigenvalue, just as the harmonics $\ell=0$ and $\ell=1$. The spectrum of the thermodynamic Hessian \eqref{thermoHessian} is generalized by the spectrum of the Euclidean action for (stationary and axisymmetric) off-shell perturbations of the black hole. This spectrum determines the one-loop correction to the gravitational partition function; see \cite{Gibbons:1978ji,Gross:1982cv} for the study of these corrections. That negative eigenvalues of the thermodynamic Hessian imply negative modes of the Euclidean action has been shown in Ref.~\cite{Reall:2001ag}, in the static case, and in Refs.~\cite{Monteiro:2009tc,Dias:2010eu}, in the general case; the proof is based on a preceding  construction \cite{Whiting:1988qr,Prestidge:1999uq}. These $\ell=0$ and $\ell=1$ type modes, predicted by the Hessian \eqref{thermoHessian}, are called {\it thermodynamic negative modes}. Additional negative modes in the partition function are  {\it non-thermodynamic negative modes}, and those are the ones associated to classical instabilities of the black hole.

Ref.~\cite{Reall:2001ag} pointed out that the eigenvalue equation for the Euclidean negative modes of a vacuum black hole, which we present in the next section, is the same as the equation for the threshold stationary mode of a Gregory-Laflamme instability of a vacuum black brane, which trivially extends that black hole along extra dimensions. The negative eigenvalue of the Euclidean negative mode corresponds to (minus) the `mass-squared" coming from the dimensional reduction along the brane directions. Therefore, a local thermodynamic instability of the black hole, i.e. a negative eigenvalue of the thermodynamic Hessian, implies the existence of a classical instability of the black brane. The simplest example is the original Gregory-Laflamme instability of a Schwarzschild black brane, which is related to the negative apecific heat of the Schwarzschild black hole. This is what underlies the Gubser-Mitra conjecture \cite{Gubser:2000ec} in the vacuum case. That conjecture says that black branes which have a translational invariance are classically unstable if and only if they are locally thermodynamically unstable. However, the preceding discussion shows that the Gubser-Mitra conjecture can be refined: there should be a distinct Gregory-Laflamme type instability for each negative mode (thermodynamic or not). Therefore, we will be finding not only thermodynamic and classical instabilities of the MP black hole, but also the associated classical instabilities of the MP black branes.

The important point for our present work is that we do not expect an instability of the black hole (whose onset is a stationary mode) within the ultraspinning surface, in the MP parameter space. In the asymptocally flat case, the ultraspinning surface is defined as the locus of the first zero-mode of the thermodynamic Hessian $-S_{\alpha\beta}$, as we mentioned. However, such a zero-mode is also identified by a reduced thermodynamic Hessian,
$ H_{ij}\equiv -\lp{\frac{\partial^2{S}}{{\partial J_i}{\partial
J_j}}}\rp_M= -S_{ij}$, as shown in \cite{Dias:2010eu}.
We can thus make use of the reduced Hessian analysis for the general MP black hole done in \cite{Dias:2010maa}. 

We will consider here only perturbations preserving the equality between two of the angular momenta. The MP ultraspinning region is defined having in mind perturbations which can change the angular momenta in an arbitrary manner, i.e. considering all the eigenvalues of $H_{ij}$. Therefore, we can find a more stringent ultraspinning surface (i.e. a smaller ultraspinning region) for our perturbations if we use the angular momenta defined with respect to $\partial_{\psi_1}$ and $\partial_{\psi_2}$, so that $H_{ij}$ is $2\times 2$ matrix, rather than $3\times 3$. Specializing it to the case $d=7$, and  $a_1=a_2\equiv a$ and $a_3\equiv  b$ that interests us, we find that the parameter space locus of the ultraspinning surface is:
\begin{equation}
\frac{|a|}{r_M}=\sqrt{2\frac{r_M^2}{b^2}-1}\,. \label{eq:ultrasurf}
\end{equation}
This ultraspinning surface is the black curve in Figure~\ref{fig:phasediag}, where the parameter space of the two unequal spins sub-family is presented (ignore the coloured dots for now). Inside this surface (\ie to the left of the black curve in  Figure~\ref{fig:phasediag}), $-S_{\alpha\beta}$ has only one negative mode. There is a second eingenvalue of  $-S_{\alpha\beta}$ that is positive inside the ultraspinning surface, but then vanishes and becomes negative as the rotation increases and the ultraspinning surface is crossed (\ie when we cross the black curve in Figure~\ref{fig:phasediag} from left to right). 

The ultraspinning region is bounded by the ultraspinning surface and by the extremality curve where the black hole temperature vanishes. The parameter space locus of the extremality curve is 
\begin{equation}
\frac{|a|}{r_M}=\sqrt{2\frac{r_M^2}{b^2}+1}\,, \label{eq:ExtSurf}
\end{equation}
and is described by the blue curve in Figure~\ref{fig:phasediag}.

\setcounter{equation}{0}
\section{Ultraspinning perturbations of MP black holes}
\label{sec:PerturbationProblem}

\subsection{Perturbations and eigenvalue problem}
\label{sec:Perturbations}

We will consider perturbation modes $h_{\mu\nu}$ that preserve the $\mathbb{R} \times U(1) \times U(2)$ isometries
of the $d=7$ MP solutions with two unequal spins \eqref{eq:MPab}. This means that we will look for stationary modes which preserve the equality between two of the three angular momenta of the background geometry. The most general perturbed line element with these properties is described by the {\it ansatz}: 
\begin{eqnarray}\label{ansatz}
&& ds^2 = \frac{r^2+v^2}{X(r)}e^{2 \delta \beta}(dr-\delta\chi dv)^2+\frac{r^2+v^2}{Y(v)}e^{2 \delta \mu_1}dv^2+\frac{(r^2+a^2)(a^2-v^2)}{a^2-b^2}e^{2 \delta\mu_4}d s^2_{\mathbb{CP}^1} \nonumber\\
&&\hspace{1cm}-\frac{X(r)}{r^2+v^2}e^{2 \delta \alpha}\left[dt+\frac{a(a^2-v^2)}{a^2-b^2}e^{2 \delta\omega_3} (d \psi_1+\mathbb{A}_1)\right]^2  \nonumber\\
&&\hspace{1cm}+
\frac{Y(v)}{r^2+v^2}\frac{a^2(r^2+a^2)^2}{(a^2-b^2)^2}e^{2 \delta \mu_3}\left[(d \psi_1+\mathbb{A}_1)+\frac{a^2-b^2}{a(r^2+a^2)}e^{-2 \delta\omega_3}dt\right]^2 \nonumber\\
&&\hspace{1cm}+r^2v^2a^2b^2e^{2 \delta \mu_2}\left[\frac{d\psi_2}{a^2 b}-\frac{(r^2+a^2)(a^2-v^2)}{a(a^2-b^2)r^2v^2}e^{2 \delta\omega_1}(d \psi_1+\mathbb{A}_1)-\frac{dt}{r^2v^2}e^{2 \delta\omega_2}\right]^2,
\end{eqnarray}
where $\{ \delta\alpha, \delta\beta,\delta\chi,\delta\mu_1,\delta\mu_2,\delta\mu_3,\delta\mu_4,\delta\omega_1,\delta\omega_2,\delta\omega_3\}$
are small quantities that describe our perturbations, and they are only functions of  $(r,v)$. 

We choose to work in the traceless-transverse (TT) gauge,
\begin{equation}
h^\mu_{\phantom{\mu}\mu}=0\,, \quad \hbox{and} \quad  \nabla^\mu
h_{\mu\nu}=0\,, \label{TT}
\end{equation}
for which the linearized Einstein equations are
\begin{equation} (\triangle_L h)_{\mu\nu}\equiv - \nabla_\rho \nabla^\rho h_{\mu\nu} -2\,
R_{\mu\phantom{\rho}\nu}^{\phantom{\mu}\rho\phantom{\nu}\sigma}
h_{\rho\sigma}=0\,,
\label{Lichnerowicz}
\end{equation}
where $\triangle_L$ is the Lichnerowicz operator. Actually, following \cite{Dias:2009iu,Dias:2010eu}, we will solve the more general eigenvalue problem
\begin{equation} \label{eigenh}
(\triangle_L h)_{\mu\nu} =-k_c^2 h_{\mu\nu}\,.
\end{equation}
This problem arises when one considers the stability of the MP black string,
$ds_{\textrm{string}}^2=g_{\mu\nu}\,dx^\mu dx^\nu+dz^2$,
under Gregory-Laflamme-type perturbations, $ds_{\textrm{string}}^2\to ds_{\textrm{string}}^2+e^{\ii k_c z} h_{\mu\nu}(x)dx^\mu dx^\nu$,
where $g_{\mu\nu}$ is the metric of the MP black hole \eqref{eq:MPab}.\footnote{The label $k_c$, rather than $k$, means that this is the critical wavenumber of a Gregory-Laflamme perturbation. The perturbation decays with time for $|k|>k_c$, and grows with time for $|k|<k_c$. The stationary onset is $|k|=k_c$.}
The same problem also arises when one considers the quadratic quantum corrections to the gravitational partition function in the saddle point approximation \cite{Gross:1982cv} (see \cite{Monteiro:2009ke} for the application to the Kerr-AdS black hole, where the numerical method used here was first applied to black hole perturbations).

Equation \eqref{eigenh} describes a coupled system of partial differential equations (PDEs) that we can solve only numerically. We look for solutions of  \eqref{eigenh} instead of  \eqref{Lichnerowicz} because \textit{Mathematica} has  powerful built-in routines to solve generalized eigenvalue problems. The strategy to find zero-modes is then to look for solutions of \eqref{eigenh}, and vary the rotation parameters $(a,b)$ until we obtain $k_c =0$. An added value of this strategy is that we will be looking not only for instabilities of the MP black hole but, in addition, to instabilities of the associated black strings.   
Indeed, solutions with $k_c \neq 0$ signal new kinds of Gregory-Laflamme instabilities and inhomogeneous phases of ultraspinning black strings. On the other hand, when $k_c=0$, we will be dealing either with the thermodynamic zero-mode (whose location in parameter space -- the ultraspinning surface -- we predicted in the last section), or with the onsets of the ultraspinning instabilities of the MP black hole.

Notice that the ansatz in \eqref{ansatz} is the most general one that respects  the isometries of the background MP black hole  \eqref{eq:MPab} and is preserved under diffeormorphisms that depend only on $(r,v)$. Ultimately, this is necessary and sufficient to guarantee that \eqref{eigenh} leads to a closed system of equations.

\subsection{Boundary conditions}
\label{sec:BCs}

In this section, we will discuss in detail the boundary conditions  that we need to impose on the metric perturbations in
order to solve \eqref{eigenh}. In the present situation, we have to specify boundary conditions at the horizon, $r=r_+$, at asymptotic infinity, $r\to\infty$, and at $v=|a|$ and $v=|b|$.

The metric perturbations must be regular on the future event horizon ${\cal H}^+$. To find the associated boundary conditions we first introduce the ingoing Eddington-Finkelstein coordinates $(V,\widetilde{\psi}_1,\widetilde{\psi}_2)$:
\begin{equation}
 dt = dV-\frac{r^2+a^2}{X(r)}\,dr\,,\qquad  d\psi_1=d\widetilde{\psi}_1 +\frac{a^2-b^2}{a \,X(r)}\,dr\,,\qquad  d\psi_2=d\widetilde{\psi}_2  -\frac{b(r^2+a^2)}{r^2X(r)}\,dr\,. 
\end{equation}
In these coordinates, the background geometry  \eqref{eq:MPab} is manifestly regular on ${\cal H}^+$. Regularity of the perturbed geometry is guaranteed if the components $h_{\mu\nu}$ of the perturbation in the Eddington-Finkelstein coordinates are regular. Translated to the perturbation notation introduced in the {\it ansatz} \eqref{ansatz}, the boundary conditions at the horizon are then 
\begin{equation}
\delta\alpha,\,\delta\beta,\,\delta\mu_1,\,\delta\mu_2,,\,\delta\mu_3,\,\delta\mu_4,\,\delta\omega_1,\,\delta\omega_2 =\mathcal{O}(1)\,, \qquad \delta\omega_3,\,\delta\chi =\mathcal{O}(r-r_+)\,.
\label{eq:BC:H}
\end{equation}
Quite importantly, note that perturbations obeying these boundary conditions preserve the temperature and the angular velocities of the background black hole. This is the case because we have imposed the regularity of the perturbation $h_{\mu\nu} dx^\mu dx^\nu$ separately, i.e. seen as a symmetric 2-tensor on the unperturbed background, rather than imposing only the regularity of the perturbed geometry. 

At spatial infinity, $r \to \infty$, we are interested in boundary conditions that preserve the asymptotic flatness of the spacetime. 
In this asymptotic region, the Lichnerowicz eigenvalue equations \eqref{Lichnerowicz} reduce simply to $\square h_{\mu\nu} \simeq k_c^2 h_{\mu\nu}$. Its regular solutions decay as
\begin{equation}
\label{eqn:BC:infinity}
h_{\mu\nu}{\bigr|}_{r\rightarrow\infty}\sim \frac {1}{r^\alpha}\,e^{-k_c\,r} \rightarrow 0 \,,
\end{equation}
for some constant $\alpha \geq 0$ that depends on the particular metric component. Therefore, for $k_c\neq 0$, asymptotic flatness is preserved since the perturbations vanish exponentially. (Note that, in our numerical method, we find modes that approach as close as desired  $k_c=0$, but without ever reaching it exactly.)

The strategy to discuss the boundary conditions at $v=|a|$ and $v=|b|$ is similar to the one used for the event horizon. We first zoom the background geometry near these critical points in coordinates where the geometry is manifestly regular. The boundary conditions for the metric perturbations can then be determined by demanding that the components $h_{\mu\nu}$ are regular in the above coordinates where the background metric is explicitly regular. We leave the details of this analysis to Appendix~\ref{app:BCpoles}, and quote the final conclusion here. 

The boundary conditions at $v=|a|$ are
\begin{equation}
\delta\alpha,\,\delta\beta,\,\delta\mu_1,\,\delta\mu_2,\,\delta\omega_1,\,\delta\omega_2,\,\delta\omega_3,\,\delta\chi =\mathcal{O}(1)\,, \qquad \delta\mu_3-\delta\mu_4,\;\delta\mu_4-\delta\mu_1 =\mathcal{O}(v-a)\,.
\label{eq:BC:va}
\end{equation}
On the other hand, the boundary conditions at $v=| b|$ are
\begin{eqnarray}
&& \delta\alpha,\,\delta\beta,\,\delta\mu_1,\,\delta\mu_2,\,\delta\mu_4,\,\delta\omega_2,\,\delta\chi =\mathcal{O}(1)\,,\\
&&  \delta\mu_3-\delta\mu_1=\mathcal{O}(v-b)\,, \qquad \delta\omega_1-\frac{a^2}{r^2+a^2}\,\delta\omega_2=\mathcal{O}(v-b)\,\qquad \delta\omega_3=\mathcal{O}(v-b)\,.\nonumber
\label{eq:BC:vb}
\end{eqnarray}

As a verification of our regularity analysis, we have checked that the boundary conditions  \eqref{eq:BC:H}-\eqref{eq:BC:vb} are consistent both with the Lichnerowicz eigenvalue equations \eqref{Lichnerowicz} and with the TT gauge conditions \eqref{TT}. Indeed, the first term in the series expansion of the eigenvalue Lichnerowicz equations vanishes after we impose the aforementioned boundary conditions. Moreover, the first term of a series expansion of the TT gauge conditions is also consistent with \eqref{eq:BC:H}-\eqref{eq:BC:vb}.

\subsection{Strategy to find the unstable modes}
\label{sec:Strategy}

The strategy to solve the  Lichnerowicz eigenvalue problem \eqref{eigenh} for  the ten metric perturbations described in \eqref{ansatz}, namely $\{ \delta\alpha, \delta\beta,\delta\chi,\delta\mu_1,\delta\mu_2,\delta\mu_3,\delta\mu_4,\delta\omega_1,\delta\omega_2,\delta\omega_3\}$, subject to the TT gauge conditions \eqref{TT} is as follows.

We first introduce the orthogonal tetrad basis $\{  e^{(a)}\}$:
\begin{eqnarray}\label{eq:tetrad}
&&\hspace{-0.6cm} e^{(0)}=dt+\frac{a\left(a^2-v^2\right)}{a^2-b^2}\left( d\psi_1+\mathbb{A}_1\right), \quad e^{(1)}=dr, \quad e^{(2)}=dv,  \quad e^{(3)}=dt+\frac{a\left(r^2+a^2\right)}{a^2-b^2}\left(d\psi_1+\mathbb{A}_1\right), \nonumber\\
&&\hspace{-0.6cm} e^{(5)}=r v a b\left(-\frac{1}{r^2v^2}\,dt-\frac{\left(r^2+a^2\right)\left(a^2-v^2\right)}{a\left(a^2-b^2\right)r^2v^2}\left(d\psi_1+\mathbb{A}_1\right)+\frac{1}{a^2b}\,d\psi_2\right),  \quad e^{(i)}=\widehat{e}^{(i)}{}_{\hat{a}} dx^{\hat{a}}\,,
\end{eqnarray}
where $\widehat{e}^{(i)}$ is the vielbein of the $\mathbb{CP}^1\cong S^2$ manifold. In this particular tetrad basis, the  Lichnerowicz eingenvalue equations simplify considerably.

The map between the metric perturbations  $h_{\mu\nu}$ in the coordinate basis and the perturbations $h_{ab}$ in the tetrad basis follows straightforwardly from $h_{ab}=e_{(a)}^{\phantom{\mu}\phantom{\mu}\mu}e_{(b)}^{\phantom{\mu}\phantom{\mu}\nu}h_{\mu\nu}$. The ten metric perturbations $\{ \delta\alpha, \delta\beta,\delta\chi,\delta\mu_1,\delta\mu_2,\delta\mu_3,\delta\mu_4,\delta\omega_1,\delta\omega_2,\delta\omega_3\}$ described in the {\it ansatz} \eqref{ansatz}  
then map to the ten tetrad components $\{ h_{00},h_{02},h_{03},h_{11},h_{12},h_{22},h_{33},h_{34},h_{44},h_{55}\}$.

The TT gauge conditions \eqref{TT} allow us to eliminate $\{ h_{00},h_{03},h_{34} \}$ in terms of the other seven functions $\{h_{02},h_{11},h_{12},h_{22},h_{33},h_{44},h_{55}\}$ and their first derivatives.  Making these substitutions in the full set of the perturbation equations \eqref{eigenh}, we find that only seven equations remain of second order. Explicitly, these equations are
\begin{eqnarray}
&& \left(\Delta _Lh\right)_{02}=-k_c^2h_{02}\,,\quad \left(\Delta _Lh\right)_{11}=-k_c^2h_{11}\,,\quad  \left(\Delta _Lh\right)_{12}=-k_c^2h_{12}\,,\quad 
\left(\Delta _Lh\right)_{22}=-k_c^2h_{22}\,,\nonumber \\
&&\left(\Delta _Lh\right)_{33}=-k_c^2h_{33}\,,\quad  \left(\Delta _Lh\right)_{44}=-k_c^2h_{44}\,,\quad  \left(\Delta _Lh\right)_{55}=-k_c^2h_{55}\,,
\label{eqn:finalsystem}
\end{eqnarray}
and they constitute our final set of equations to be solved.
A non-trivial consistency check of our procedure is to verify that the final equations \eqref{eqn:finalsystem} imply that the three remaining equations, which are of third order, are automatically satisfied. 

We use spectral methods to solve numerically the final eigenvalue problem \eqref{eqn:finalsystem}. Less computational power is required in the implementation of the spectral method if all functions obey Dirichlet boundary conditions on all boundaries. Thus, we introduce new independent functions $\{q_i\}$ related to the independent metric functions by
\begin{equation}
\label{eq:qs}
\begin{aligned}
&q_1(r,v)=\frac{1}{r_+^2}(v-a)(v-b)\left(1-\frac{r_+}{r}\right){}^2h_{11}(r,v)\,,\quad q_2(r,v)=\frac{1}{r_+^2}(v-a)(v-b)\left(1-\frac{r_+}{r}\right)h_{12}(r,v)\,,\\
& q_3(r,v)=\frac{1}{r_+^4}(v-a)^2(v-b)^2\left(1-\frac{r_+}{r}\right)h_{22}(r,v)\,,\\
& q_4(r,v)=r_+^2\left(1-\frac{r_+}{r}\right)\frac{\left(r^2+v^2\right)^2 h_{33}(r,v)-Y(v)^2 h_{22}(r,v)}{\left(r^2+v^2\right)^2Y(v)}\,,\\
& q_5(r,v)=\frac{1}{r_+^2}(v-a)(v-b)\left(1-\frac{r_+}{r}\right)h_{44}(r,v)\,,\\
& q_6(r,v)=\frac{1}{r_+^2}(v-b)^2\left(1-\frac{r_+}{r}\right)\frac{\left(a^2-b^2\right)^2h_{55}(r,v)-a^2Y(v)^2h_{22}(r,v)}{\left(a^2-b^2\right)^2 Y(v)}\,,\qquad q_7(r,v)=h_{02}(r,v) \,. \\
\end{aligned}
\end{equation}
The Dirichlet boundary conditions for the $q_i$'s are equivalent to the boundary conditions \eqref{eq:BC:H},  \eqref{eqn:BC:infinity}, \eqref{eq:BC:va} and \eqref{eq:BC:vb}.\footnote{The reader might ask the reason for the particular combinations in the definitions of $q_4$ and $q_6$, in contrast with the simpler definitions of the other $q_i$'s. These combinations maximize the information on the boundary conditions. To illustrate this take \eqref{eq:BC:vb}. One has $\delta\mu_3-\delta\mu_1=\mathcal{O}(v-b)$. So our fundamental variables should be $\{\delta\mu_1,\, \delta\mu_3-\delta\mu_1\}$,   instead of considering only the leading behaviour of $\{\delta\mu_1,\,\delta\mu_3\}$; otherwise the information that $\delta\mu_3-\delta\mu_1=0$ is lost. A similar reasoning applies to the variables involved in the last relation of \eqref{eq:BC:va}. The combinations chosen for the definition of $q_4$ and $q_6$ account for this issue, in the notation of the tetrad metric components.}
 It is also convenient to work with coordinates that range in a unit interval. We thus introduce the new coordinates
\beq
y=1-\frac{r_+}{r}\,,\qquad x=\frac{\alpha-v}{\alpha-\beta}\,, \qquad \hbox{with} \quad \alpha\equiv \hbox{min}\{|a|,|b|\} \,\,\,\, \hbox{and} \,\,\,\, \beta\equiv \hbox{max}\{|a|,|b|\}\,, 
\eeq
so that $0\leq y \leq 1$ and $0\leq x\leq 1$.

\setcounter{equation}{0}
\section{Results and discussion}
\label{sec:results}

\begin{figure}[t]
\centering
\includegraphics[width =8.0 cm]{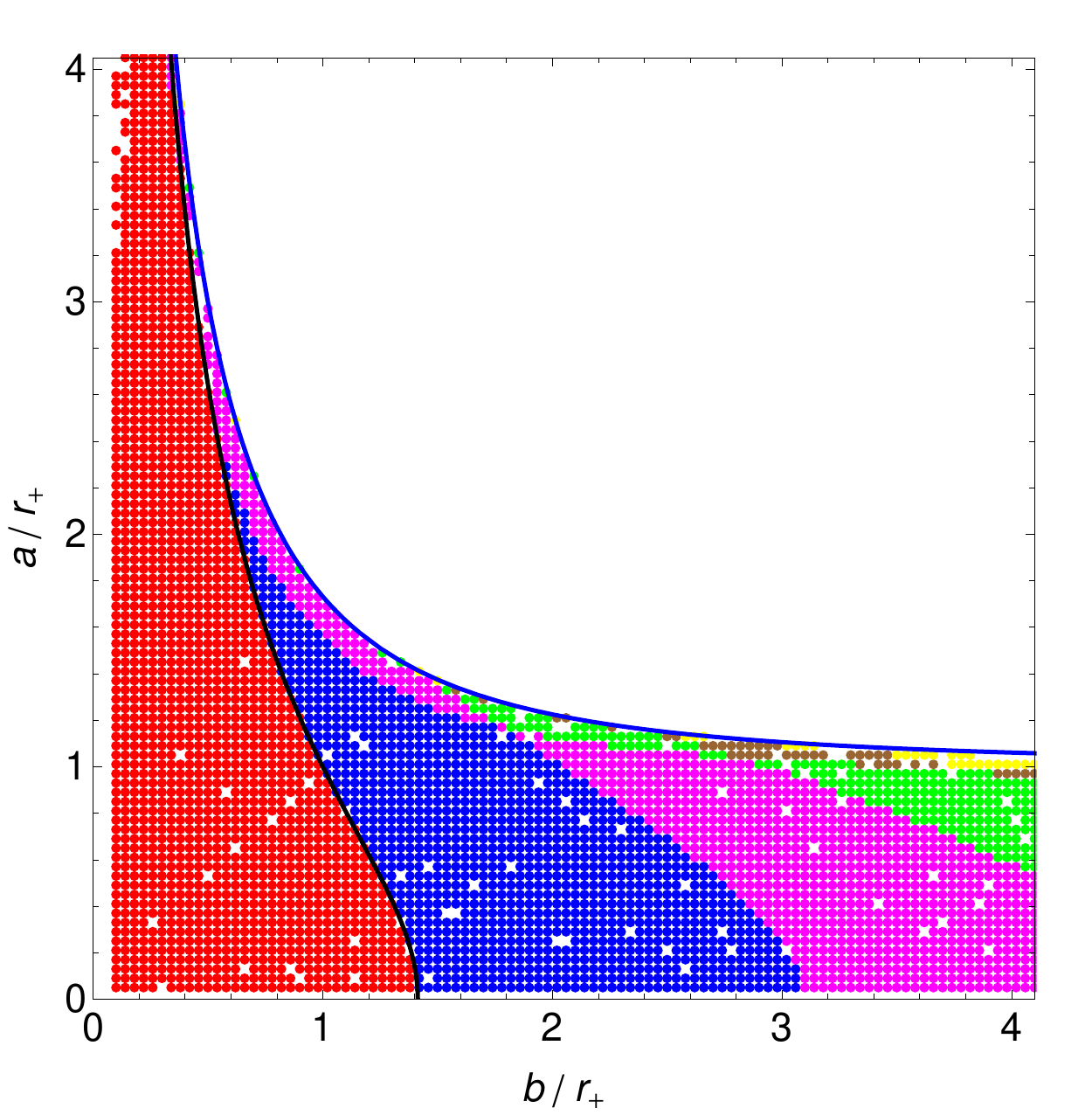}
\caption{\label{fig:phasediag} Phase diagram of MP solutions with two unequal spins in $d=7$. The axes are the rotation parameters $a$ and $b$, normalized to the horizon radius $r_+$. The colour code is the number of negative modes found numerically: red 1, blue 2, violet 3, green 4, brown 5, yellow 6. The blue line represents extremality, the black line between red and blue represents the ultraspinning surface, and the line between the blue and violet regions is the onset of the ultraspinning instability. The scattered white dots are numerical faults.}
\end{figure}

We determined numerically the negative modes in the Lichnerowicz eigenvalue problem \eqref{eigenh} for the MP sub-family with two unequal spins in $d=7$. Our perturbations $h_{\mu\nu}$ are stationary and preserve the $U(1)\times U(2)$ spatial isometry of the background solution, which implies that they preserve the equality between two of the three MP spins. In Figure~\ref{fig:phasediag}, we present the number of negative modes for a grid of points on the parameter space. The axes represent the rotation parameters, $a$ and $b$, normalized to the horizon radius $r_+$. The blue line that binds the parameter space represents extremality. The colour coding is the following: red denotes the existence of 1 negative mode, blue 2, violet 3, green 4, brown 5, and yellow 6.

For small rotation parameters, in the red region, there is a single negative mode. This is the negative mode that any asymptotically flat vacuum black hole possesses \cite{Dias:2010eu}; for $a=b=0$, it is the negative mode of the Schwarzschild black hole \cite{Gross:1982cv}, which is associated with its negative specific heat. In the blue region, an additional negative mode exists. This negative mode is also thermodynamic in nature. It appears along the black line, which represents the ultraspinning surface, discussed in Section~\ref{subsec:ultrathermo}. This surface, given by the expression \eqref{eq:ultrasurf}, is defined by the appearance of a second negative eigenvalue of the thermodynamic Hessian \eqref{thermoHessian}. Therefore, the fact that it precisely separates the red and blue regions indicates an excelent agreement between our expectations and the numerical results.

The most interesting line in Figure~\ref{fig:phasediag} is the separation between the blue and the violet regions.\footnote{ The reason for choosing the horizon radius as a normalization is that the ``distance" between the blue region and extremality is larger, say for $a=b$. Had we normalized the grid with respect to the mass-radius $r_M$, it would be impossible to see the region of interest for equal spins without zooming in substantially. On the other hand, this normalization makes the parameter space seem unbounded for $b=0$, because $r_+\to0$ in the singular extremal limit.}
Across that line, a third negative mode appears, one which is not related to the thermodynamic Hessian \eqref{thermoHessian}. This mode is the first stationary onset of the ultraspinning instability (within our symmetry restrictions of keeping two unequal spins), and it marks also a bifurcation to a new family of black holes. The prediction of the ultraspinning conjecture is verified, as this line only appears for rotation larger than the ultraspinning surface. The lines between the regions violet/green, green/brown and brown/yellow (and others which would appear had we zoomed in the appropriate region, or extended our analysis to higher $b/r_+$) signal new instabilities/bifurcations. Let us also point out that each of those lines (and also the black one, the ultraspinning surface) corresponds to the appearance of a new Gregory-Laflamme-type instability of the MP black branes.

One of the main results of this paper is that the blue/violet line, i.e. the onset of the first ultraspinning instability, continuously connects the singly-spinning case ($a=0$) and the equal spins case ($a=b$). These sectors have been studied before in Refs.~\cite{Dias:2009iu} and \cite{Dias:2010eu,Durkee:2010ea}, respectively, and the numerics matches precisely with the previous results. The question which remained was whether the instabilities in the singly-spinning sector and in the equal spins sector were of the same nature. We recall that there is a geometric argument that explains the instability in the singly-spinning case \cite{Emparan:2003sy}, namely that for high rotation the black hole resembles locally a black brane, which suffers from the Gregory-Laflamme instability. However, no analogous argument has been proposed for the instability in the equal spinning sector, for which the rotation is bounded by extremality. We have shown here that the instability in the equal spins sector is indeed of the same nature as the originally proposed instability in the singly-spinning sector. Their onsets are continuously connected in the MP parameter space.\footnote{Note that Refs.~\cite{Dias:2010eu,Durkee:2010ea} consider general axisymmetric perturbations in the equal spins sector, without our symmetry restriction of keeping the spatial isometry $U(1)\times U(2)$. The onset of the ultraspinning instability found here is a single perturbation, while in those works there was a large degeneracy of the harmonics of $\mathbb{CP}^2$ from which the perturbations were constructed. This means that several ``onset surfaces'' will meet at the equal spins point in the full MP parameter space, and we have studied only one of them, which was sufficient for our purposes.}

What is possibly our main result is that the blue/violet onset line seems to occur everywhere before extremality. This would mean that we can always find unstable black holes in a small enough neighbourhood of extremality, and that the extremal black holes are themselves unstable. This raises several questions.

We can follow numerically only a finite number of onset lines, e.g. blue/violet, violet/green, green/brown, etc.. There are then three possibilities which are difficult to check numerically. The first is that all the onset lines squeeze in between the ultraspinning surface and extremality, beeing closer to extremality for higher number of associated negative modes. This would mean that the extremal black holes would be severely unstable solutions, since they would suffer from an infinite number of distinct linear instabilities. In the second and third possibilities, there is an infinite number of onset lines which terminate at some point along the extremality line, but in a different manner. (In fact, we can't be sure that this does not occur for the blue/violet line too, for small $b/r_+$.) So the second possibility is that the onsets can occur also {\it on} the extremal line, i.e. there is a regular zero-mode for some extremal solutions, which would connect them to new extremal black holes. This is not possible for perturbations of a scalar field on an extremal black hole background, since the scalar field cannot be regular on the near-horizon geometry \cite{FernandezGracia:2009em}. The third possibility is that the same is true for gravitational perturbations, in which case the onset lines terminate at extremality but the zero-mode is not regular for the extremal solution, so that the bifurcation does not extend to extremality. For instance, in the scalar field condensation analysis of Ref.~\cite{Dias:2010ma}, the zero-mode line touches the extremality line tangentially, but at that point one does not have a hairy extremal black hole.

The most relevant question raised by the instability of extremal solutions, in the MP sub-family studied here, is whether this extends to MP extremal solutions beyond this sub-family. Recall that, in $d=5$, no instability was found on the equal spins sector \cite{Murata:2008yx}. However, only in $d\geq6$ are there regimes with unbounded angular momenta, so the parameter space is qualitatively different. So the question is: are there any extremal Myers-Perry black holes which are stable in $d\geq6$? 

This is the first work which studies the stability of MP solutions with two independent angular momenta turned on. As we emphasized, we focus on axisymmetric perturbations. It would be interesting to know how perturbations breaking that symmetry, such as the ones studied in Refs.~\cite{Shibata:2009ad,Shibata:2010wz}, constrain the stable parameter space. Moreover, we still lack a good understanding of the stable parameter regions of other higher-dimensional solutions, for instance black rings. The main problems in higher-dimensional vacuum black holes remain open: (i) Is a classification of higher-dimensional black holes possible? (ii) Which MP solutions are stable? (iii) Are there any stable solutions apart from ``not-so-rapidly rotating" MP black holes?

\section*{Acknowledgements}

We would like to thank Troels Harmark, Niels Obers and Harvey Reall for discussions. OJCD acknowledges financial support provided by the European Community through the Intra-European Marie Curie contract PIEF-GA-2008-220197. RM is supported by the Danish Council for Independent Research - Natural Sciences (FNU).

\appendix

\setcounter{equation}{0}
\section{Boundary conditions at $v=|a|$ and $v=|b|$}
\label{app:BCpoles}

In this Appendix, we give the details of the regularity analysis at $v=a$ and $v=b$ (take $a,b\geq0$ for simplicity), which leads to the boundary conditions \eqref{eq:BC:va} and \eqref{eq:BC:vb}.

Let us start with the analysis of $v=a$. For $v\approx a$, we can write $Y(v)=Y'(a)(v-a)+O[(v-a)^2]$ with $Y'(a)=2(b^2-a^2)/a$. To zoom the geometry in this region, we introduce the new variable $\rho$: $v=a+\rho^2/4$. We introduce also the manifestly regular 1-forms $E^\rho=\rho d\rho$ and $E^{\psi_1}=\rho^2(d\psi_1+\mathbb{A}_1)$. The geometry near $v=a$ then reads
\begin{eqnarray}\label{eq:MPnearVa}
&& \hspace{-1cm}ds^2 \approx \frac{r^2+a^2}{X(r)}\,dr^2 +\frac{a(r^2+a^2)}{2(b^2-a^2)}\left( ds^2_{\textrm{ball}_4}-\frac{2(b^2-a^2)}{a(r^2+a^2)}\,E^{\psi_1}dt+ \rho^2 \,\frac{(b^2-a^2)^2}{a^2(r^2+a^2)^2} dt^2    \right)\\
&& \hspace{0.0cm}  - \frac{X(r)}{r^2+a^2} \left(  dt+ \frac{a}{2(b^2-a^2)} \,E^{\psi_1} \right)^2
+r^2a^4b^2\left( \frac{d\psi_2}{a^2b} - \frac{r^2+a^2}{2r^2a^2(b^2-a^2)}\,E^{\psi_1}  -\frac{dt}{r^2+a^2} \right)^2,\nonumber
\end{eqnarray}
where we used $ds^2_{\textrm{ball}_4} = d\rho^2+\rho^2 [(d\psi_1+\mathbb{A}_1)^2+ds^2_{\mathbb{CP}^1}]$. This  line element is manifestly regular as $\rho\to 0$. The boundary conditions for the metric perturbations can now be determined by demanding that $h_{\mu\nu} dx^\mu dx^\nu$ is a regular symmetric 2-tensor, i.e. that it has regular components when expressed in the above coordinates, where the background metric is regular. Near $\rho=0$ ($v=a$), the perturbation reads
\begin{equation}\label{eq:PertnearVa}
\begin{aligned}
\hspace{-0.1cm}h_{\mu\nu}\,dx^\mu\,dx^\nu &\approx  \frac{r^2+a^2}{X(r)}\left(2\delta\beta\,dr^2 -\delta\chi E^\rho dr \right) + \frac{a(r^2+a^2)}{b^2-a^2}
{\biggl [} \left(\delta\mu_1 \,d\rho^2 +  \delta\mu_3\,\rho^2 (d\psi_1+\mathbb{A}_1)^2 +\delta\mu_4 \,ds^2_{\mathbb{CP}^1}\right) \\
&\hspace{-1cm}~+\delta\mu_3\left(-2 E^{\psi_1} dt +\frac{b^2-a^2}{a(r^2+a^2)}\,\rho^2dt^2\right) {\biggr ]}
 +2r^2a^4b^2 \delta\mu_2\left(\frac{d\psi_2}{a^2b}  + \frac{r^2+a^2}{2r^2a^2(b^2-a^2)}\,E^{\psi_1}  -\frac{dt}{r^2a^2}\right)^2 \\
&\hspace{-1cm}~ +\left(\frac{2ab^2(r^2+a^2)}{b^2-a^2}\,\delta\omega_1 \,E^{\psi_1}-4a^2b^2 \,\delta\omega_2\,dt  \right)  \left(  \frac{d\psi_2}{a^2b} + \frac{r^2+a^2}{2r^2a^2(b^2-a^2)}\,E^{\psi_1}  -\frac{dt}{r^2a^2}\right)^2\\
&\hspace{-1cm}~
 + \frac{2a X(r)}{r^2+a^2}\frac{E^{\psi_1}}{b^2-a^2} \left(  dt+ \frac{a^2}{2(b^2-a^2)} \,E^{\psi_1} \right)\delta\omega_3+2dt\left( E^{\psi_1}-\frac{b^2-a^2}{a(r^2+a^2)}\,\rho^2 dt \right)\delta\omega_3\\
&\hspace{-1cm}~ -\frac{2 X(r)}{r^2+a^2}\,\delta\alpha\left( dt+\frac{a^2}{2(b^2-a^2)}\,E^{\psi_1}\right)^2.
\end{aligned}
\end{equation}
Regularity of the perturbed geometry as $v\to a$ then requires the boundary conditions \eqref{eq:BC:va}.
Note that the last condition in  \eqref{eq:BC:va} guarantees regularity of the last contribution in between curved brackets in the first line of \eqref{eq:PertnearVa}; see the contribution $ds^2_{\textrm{ball}_4}$ in the background geometry \eqref{eq:MPnearVa}. 

Finally, consider $v=b$.  In the neighborhood of $v=b$, we can write $Y(v)=Y'(b)(v-b)+O[(v-b)^2]$ with $Y'(b)=2(a^2-b^2)/b$. The new variable $\widetilde{\rho}$, defined through $v=b-\widetilde{\rho}^{\,2}/4$,  allows us to zoom the geometry in this region. Introduce the new coordinates 
$d\widetilde{t} = dt +  b \,d\psi_2$ and $d\widetilde{\psi}_1 = d\psi_1 - \frac{b}{a}\,d\psi_2$. Consider also the regular 1-forms  $E^{\widetilde{\rho}}=\widetilde{\rho}\, d\widetilde{\rho}$ and $E^{\psi_2}=\widetilde{\rho}^2 d\psi_2$. The background geometry, in the vicinity of $v=b$, is then
\begin{eqnarray} \label{eq:MPnearVb}
&& \hspace{-1cm} ds^2 \approx \frac{r^2+b^2}{X(r)}\,dr^2 +\frac{b(r^2+b^2)}{2(b^2-a^2)}{\biggl [} ds^2_{\textrm{ball}_2}
+\frac{a^2}{b^2}\frac{(r^2+a^2)^2}{(r^2+b^2)^2}\,\widetilde{\rho}^{\,2}\left(  (d\widetilde{\psi}_1+\mathbb{A}_1)-\frac{b^2-a^2}{a(r^2+a^2) }d\widetilde{t} \right)^2  \nonumber \\
&& +2\frac{a}{b}\frac{r^2+a^2}{r^2+b^2}\left(  (d\widetilde{\psi}_1+\mathbb{A}_1)-\frac{b^2-a^2}{a(r^2+a^2) }d\widetilde{t} \right)\,E^{\psi_2} {\biggr ]}+(r^2+a^2)ds^2_{\mathbb{CP}^1} \nonumber \\
&& +r^2\left(  (d\widetilde{\psi}_1+\mathbb{A}_1)-\frac{a}{r^2}d\widetilde{t} \right)^2 -\frac{X(r)}{r^2+b^2}\left(  d\widetilde{t} +a(d\widetilde{\psi}_1+\mathbb{A}_1)\right)^2, 
\end{eqnarray}
where we used $ds^2_{\textrm{ball}_2} = d\widetilde{\rho}^2+\widetilde{\rho}^2 d\psi_2^2$. This is a manifestly regular geometry as $\widetilde{\rho}\to 0$. Requiring that the components $h_{\mu\nu}$ are regular when expressed in the these coordinates fixes the boundary conditions at $v=b$ ($\widetilde{\rho}=0$). In the neighborhood of $v=b$, the perturbation reads
\begin{equation}\label{eq:PertnearVb}
\begin{aligned}
\hspace{-0.1cm}h_{\mu\nu}\,dx^\mu\,dx^\nu &\approx 
\frac{(r^2+b^2)b}{2(b^2-a^2)}{\biggl [} \delta \mu_1\, d\widetilde{\rho}^2+ \delta \mu_3 \, \widetilde{\rho}^{\,2}\left( d\psi_2 +\frac{a}{b}\frac{r^2+a^2}{r^2+b^2}  (d\widetilde{\psi}_1+\mathbb{A}_1) -\frac{b^2-a^2}{b(r^2+b^2)}d\widetilde{t} \right)^2 {\biggr ]} \\
&\hspace{-1cm}~-\frac{4}{r^2}\left[ (r^2+a^2)(d\widetilde{\psi}_1+\mathbb{A}_1) -a \,d\widetilde{t} \,\right]\left[  (r^2+a^2) \delta\omega_1 \left( \frac{b}{a}d\psi_2 +d\widetilde{\psi}_1+\mathbb{A}_1 \right)-a  \delta\omega_2 \left( b\,d\psi_2 - d\widetilde{t} \right) \right]
 \\
&\hspace{-1cm}~-\delta\omega_3\,\frac{4aX(r)}{r^2+b^2}\left( \frac{b}{a}d\psi_2 +d\widetilde{\psi}_1+\mathbb{A}_1 \right)\left[ d\widetilde{t} +a(d\widetilde{\psi}_1+\mathbb{A}_1)\right]+2\delta\mu_2 \,\left[ (r^2+a^2)(d\widetilde{\psi}_1+\mathbb{A}_1)-a \,d\widetilde{t}\, \right]^2\\
&\hspace{-1cm}~-\widetilde{\rho}^2 \delta\omega_3\,\frac{2(r^2+a^2)}{b^2r^2(r^2+b^2)}\left( b \,d\psi_2 - d\widetilde{t}\, \right)\left[ (r^2+a^2)(d\widetilde{\psi}_1+\mathbb{A}_1)-a\, d\widetilde{t} \,\right]\\
&\hspace{-1cm}~
 +2\frac{r^2+b^2}{X(r)}\left(\delta\beta\,dr^2 +\delta\chi \,dr dv \right)   +2(r^2+a^2)\delta\mu_4 \,ds^2_{\mathbb{CP}^1} - \delta\alpha\,\frac{2X(r)}{r^2+b^2}\left[ d\widetilde{t}+ a(d\widetilde{\psi}_1+\mathbb{A}_1)  \right]^2.
\end{aligned}
\end{equation}
It follows that the perturbed geometry is regular as $v\to b$ if and only if the boundary conditions \eqref{eq:BC:vb} are obeyed.
Note that in the second line of \eqref{eq:BC:vb}, the first, second and third conditions guarantee the regularity of the first, second and third lines of \eqref{eq:PertnearVb}, respectively. 


\end{document}